# Phonon-assisted transport and hole–phonon coupling in GaAs double quantum dots

Jin Leng(冷进),[1,2,3,#] Wei-Zhu Liao(廖伟筑),[1,2,3,#] Hao-Tian Jiang(姜皓天),[1,2,3] Di Liu(刘頔),[4] Bao-Chuan Wang(王保传),[4] Gang Cao(曹刚),[1,2,3,4] Hai-Ou Li(李海欧),[1,2,3,4,*] and Guo-Ping Guo(郭国平)[1,2,3,4]

[1] *Laboratory of Quantum Information, University of Science and Technology of China, Hefei, Anhui 230026, China*

[2] *Anhui Province Key Laboratory of Quantum Network, University of Science and Technology of China, Anhui 230026, China*

[3] *CAS Center for Excellence and Synergetic Innovation Center in Quantum Information and Quantum Physics, University of Science and Technology of China, Hefei, Anhui 230026, China*

[4] *Hefei National Laboratory, University of Science and Technology of China, Hefei, 230088, China*

[#] These authors contributed equally to this work

[*] Corresponding author: haiouli@ustc.edu.cn;

Hole–phonon interactions play an important role in transport and decoherence processes in semiconductor quantum dots. Here we investigate hole–phonon coupling in a gate-defined GaAs double quantum dot integrated with a quantum point contact charge sensor. Under finite source–drain bias, pronounced oscillatory stripe patterns appear near specific charge transition regions in the charge stability diagram. We attribute these oscillations to phonon emission during inelastic interdot tunneling. A theoretical model including piezoelectric hole–phonon coupling reproduces the observed patterns. Furthermore, our analysis shows that the oscillations emerge only in particular charge configurations. Our results provide direct insight into phonon-assisted transport and coherent hole–phonon interactions in semiconductor quantum dots.

*1. Introduction.* Cavity quantum electrodynamics has long provided a powerful framework for studying the interaction between discrete quantum systems and bosonic fields, most prominently atoms coupled to photons inside optical or microwave cavities[1]. In solid-state systems, semiconductor quantum dots (QDs) have become versatile platforms for exploring analogous light–matter interactions[2-6], enabling the investigation of phenomena such as vacuum Rabi splitting[7], nonclassical states[8], and coherent spin–photon coupling[9-11].

Beyond photon-mediated interactions, coupling between charge carriers and lattice vibrations (phonons) plays a fundamental role in semiconductor nanostructures[12-13]. Phonons act both as an intrinsic environment responsible for dissipation and decoherence[14-15], and as active participants in transport processes through energy exchange with confined electrons or holes[16]. Owing to the relatively low sound velocity in solids, acoustic phonons possess wavelengths that can be comparable to or smaller than the characteristic length scales of quantum dot devices[17], making phonon-induced effects particularly prominent[18].

In QDs, the discrete energy spectrum and controllable tunneling processes provide a natural platform for probing carrier–phonon interactions through transport measurements[19]. In particular, inelastic interdot tunneling accompanied by spontaneous phonon emission has been extensively studied in AlGaAs heterostructures[20], carbon nanotubes[21-22], graphene[23], and semiconductor nanowires[24-25], and has been established as a key mechanism for energy relaxation in coupled quantum-dot systems[19-20]. In recent years, studies have shown that phonon interactions can significantly influence the fidelity of spin qubits[26], while other works have explored the possibility of realizing phonon-mediated coupling between spin qubits[27]. In GaAs-based devices, piezoelectric coupling dominates the interaction between charge carriers and acoustic phonons[20], leading to particularly strong hole–phonon coupling[28] and making this system an important platform for investigating carrier–phonon interactions. Previous studies have proposed theoretical models[29-31] describing these processes and have experimentally observed quantum point contact (QPC)-induced phonon absorption[28] as well as phonon-mediated single-hole tunneling[32].

In this letter, we investigate hole–phonon interactions in a gate-defined GaAs double quantum dot (DQD) integrated with a QPC charge sensor[33-34]. By applying a finite source–drain bias across the DQD, pronounced oscillatory stripe patterns are observed near specific charge transition regions in the charge stability diagram. Through systematic measurements performed under different source–drain and QPC bias conditions, we demonstrate that these oscillations originate from phonon emission during inelastic interdot tunneling processes, where the phonon energy is supplied by the applied source–drain bias. The phonon emission process we observe is complementary to the QPC-induced phonon absorption reported previously in similar systems[28], together forming a complete picture of hole-phonon interactions. Based on a theoretical model of hole–phonon interactions[29-30] incorporating piezoelectric coupling[31-32], we numerically simulate the oscillatory stripe patterns using a master-equation approach and obtain good agreement with the experimental results. Finally, we find that the oscillatory behavior appears only in particular charge configurations, providing insight into the conditions required for phonon-assisted transport.

*2. Device and Measurement.* In this work, we fabricate a gate-defined hole DQDs on an undoped AlGaAs/GaAs heterostructure, integrated with a QPC to monitor hole tunneling events in the QDs [33-34]. The GaAs/AlGaAs heterostructure used consists of a 5 nm GaAs cap layer and a 100 nm $Al_xGa_{1-x}As$ layer grown on a 500 nm GaAs buffer. Hole (QDs are confined in the two-dimensional hole gas (2DHG) accumulated at the GaAs/AlGaAs interface. In the scanning electron microscope image, eight Ti/Au gates (4/45 nm) are deposited on the surface of the GaAs cap layer to define the confinement potential in the 2DHG, forming a DQD with an integrated QPC.A 100 nm $AlO_x$ dielectric layer is deposited on the gate structures as an insulating layer. Finally, a global Ti/Au top gate (10/70 nm) is deposited on the dielectric layer. Ohmic contacts to the 2DHG are fabricated by AuBe deposition, followed by annealing at 430 °C.

The experimental setup is shown in Fig. 1(a), where a pattern of metal electrodes is designed to provide electrostatic confinement, resulting in two separated QDs and a QPC channel in the 2DHG. A bias voltage $V_{\mathrm{sd}}$ and $V_{\mathrm{QPC}}$ are applied to the ohmic contacts of the QD and QPC, respectively. Precise control of the energy levels of the two QDs is achieved by regulating the voltages applied to gates G4 and G6 (shown in blue in the figure). The measurements were performed in a dilution refrigerator at a base temperature of about 20 mK. The strong in-plane gate-defined confinement allows for precise control of the holes in each QD. In Fig. 1(b), we plot the charging diagram at $V_{\mathrm{sd}} = 2$ mV and $V_{\mathrm{QPC}} = 0.5$ mV, showing the derivative $\mathrm{d}I_{\mathrm{QPC}}/\mathrm{d}V_{\mathrm{G6}}$ of the charge sensor current $I_{\mathrm{QPC}}$ as a function of gate voltages $V_{\mathrm{G4}}$ and $V_{\mathrm{G6}}$. This reveals the charge stability regions corresponding to different numbers $(n, m)$ of holes in each QD. We observe a series of parallel spectral lines in the figure, and we focus on the$(n+1, m)-(n, m+1)$ hole tunneling region.

As shown in Fig. 2(a), the key observation in this letter is the presence of parallel lines in this regime with an average spacing of $\delta_\varepsilon = 140\,\mu\mathrm{eV}$, obtained from the oscillation voltage spacing in the measurement through the experimentally determined level arm. The lever arm is estimated from the ratio between the source–drain bias and the detuning range of the bias triangle, yielding a value of $0.02\,\mathrm{eV/V}$. These lines indicate the state resonance of the left and right dots. The transition line between $(n+1, m)-(n, m+1)$ typically corresponds to elastic tunneling in transport measurements, where the energy levels of the left and right dots align. Excited states, several meV above the ground-state levels, cannot account for these observations in our experiments.

Based on the energy scale revealed by the spacing between the stripes, we speculate that this phenomenon is induced by phonon interactions. Fig. 1(c) shows a schematic of the energy levels of the DQDs and spontaneous emission during hole tunneling in a full quantum description. The DQD system can be modeled as a three-level system defined by the charge states of the two dots: $|0\rangle = |n_{\mathrm{L}}, n_{\mathrm{R}}\rangle$, $|R\rangle = |n_{\mathrm{L}}, n_{\mathrm{R}} + 1\rangle$ and $|L\rangle = |n_{\mathrm{L}} + 1, n_{\mathrm{R}}\rangle$.A bias voltage $\mathrm{V_{sd}} = 2$ mV is applied to the source and drain, allowing a hole to transport through the DQDs when $\mu_{\mathrm{s}} \geq \mu_{\mathrm{R}} \geq \mu_{\mathrm{L}} \geq \mu_{\mathrm{d}}$. In this tunneling process, a phonon with energy $hf = \delta$ is emitted if $\mu_{\mathrm{R}} \neq \mu_{\mathrm{L}}$, indicating that the process is inelastic tunneling. When the energy detuning $\varepsilon$ between the charge states $|L\rangle$and $|R\rangle$is varied, resonance occurs between the emitted phonon energy $\delta$ and the intrinsic phonon modes of the system[35]. This resonance enhances the inelastic tunneling process, thereby leading to the observed current oscillations.

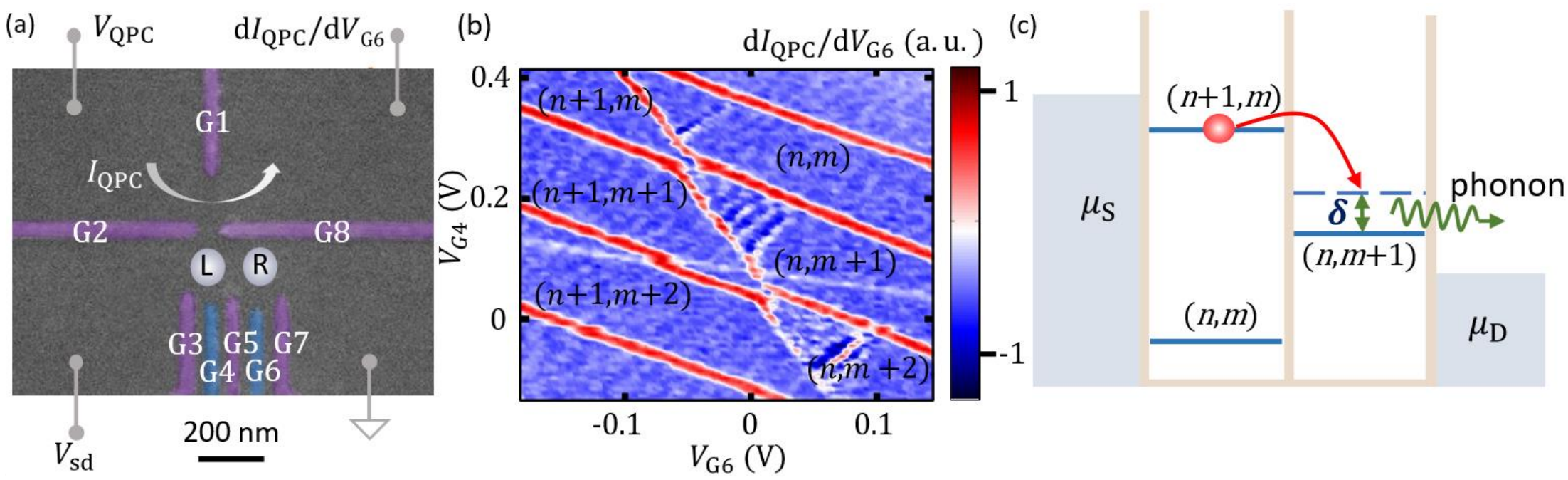

**Fig. 1.** (a) Scanning electron microscope image of the confining gates fabricated on GaAs. Seven electrodes (G2~G8) are designed to confine QDs (white circles) while a QPC channel is controlled by three gates (G1, G2 and G8). During measurement, a negative voltage $V_{\mathrm{TG}} = -2.4$ V is applied to a global top gate (invisible) to accumulate the holes at the GaAs/AlGaAs interface to form a 2DHG and direct current bias ($V_{\mathrm{sd}}$) is applied to ohmic contact to form the asymmetric potential for detecting inelastic tunneling events. (b) A wider charge configuration space is explored at a source–drain bias of $V_{sd} = 2\,\mathrm{mV}$, where the charge occupation is denoted by $(n, m)$. Clear oscillatory stripe patterns are observed in the $(n+1, m)$–$(n, m+1)$ region. In the neighboring $(n+1, m-1)$ – $(n, m)$ and $(n+1, m+1)$ – $(n, m+2)$ regions, only weak stripe features are visible, while no oscillatory behavior is observed in other

regions. (c) Energy level diagram of the DQD for positive QD bias. Inelastic tunneling between the ground states $(n+1,m)$ and $(n,m+1)$ occurs via the emission of phonons, where δ denotes the energy of the phonon. The ground state $(n,m)$ serves as the metastable state $|0\rangle$ after a hole jump out from right dot.

A similar pattern in conventional DQD systems based on a two-dimensional electron gas has been demonstrated and attributed to the back-action of the QPC, where a linear QPC bias dependence of these stripes was identified by Granger et al. [32]. We distinguish the back-action effect from our result by investigating the dependence of the stripe pattern on reservoir bias and QPC bias. In the experiment, we performed a series of measurements by varying the source–drain bias $V_{\mathrm{sd}}$ and the QPC bias $V_{\mathrm{QPC}}$, and obtained the data shown in Fig. 2(b). Here, $\varepsilon_{\max}$ denotes the maximum voltage range over which the stripe pattern is observable, as indicated in Fig. 2(a). The upper panel of Fig. 2(b) shows that $\varepsilon_{\max}$ increases monotonically as $V_{sd}$ is varied from -2 mV to 2 mV. In contrast, the lower panel indicates that $V_{\mathrm{QPC}}$ does not affect the boundary of the stripe pattern. These results indicate that the observed oscillatory stripe pattern originates from phonon emission processes driven by the energy supplied by the source–drain bias, rather than from the absorption of external phonons provided by the QPC.

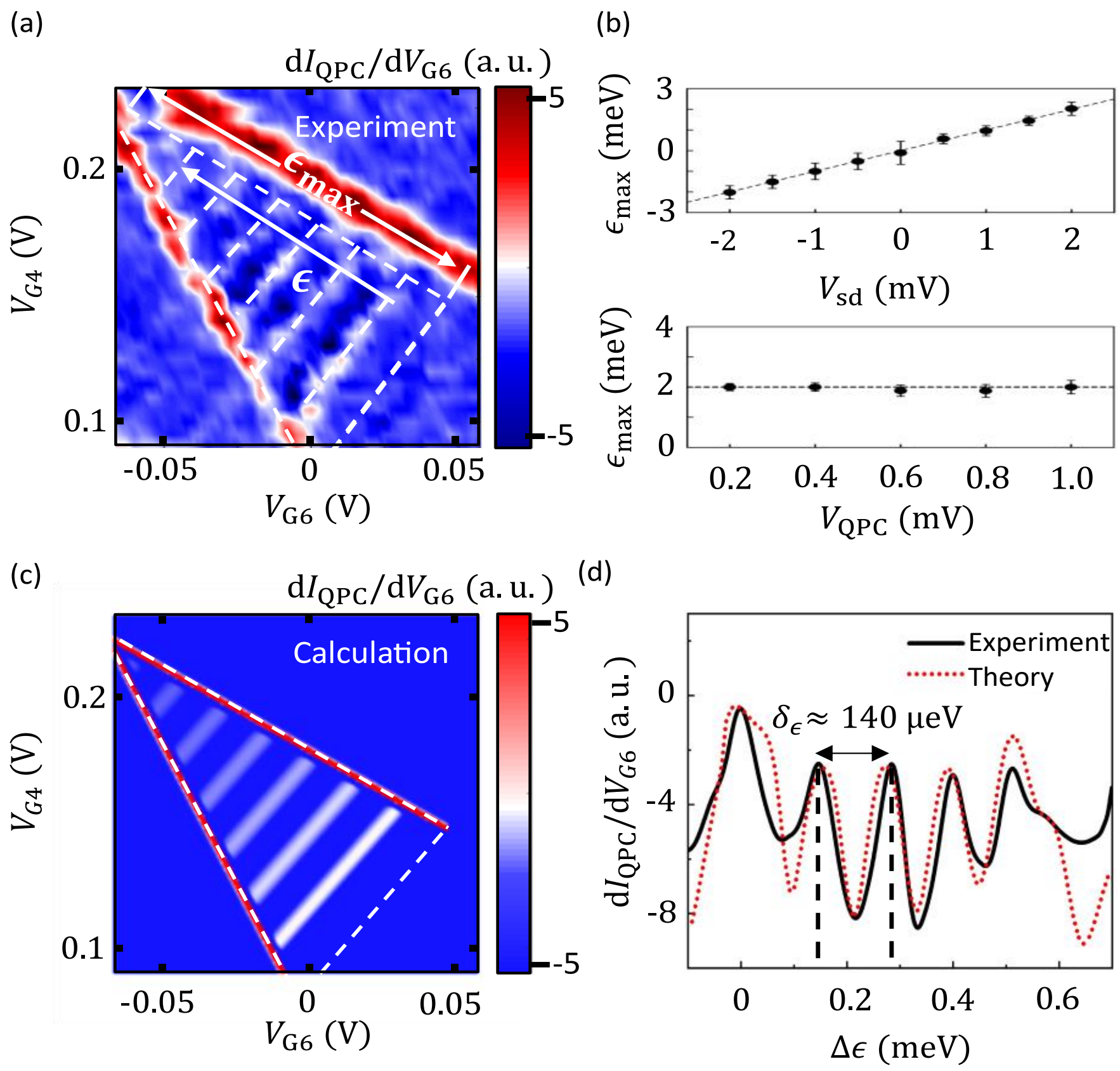

**Fig. 2.** (a) The investigated regions near the $(n+1,m)$–$(n,m+1)$ charge state under biases of $V_{sd} = 2$ mV. The white triangles mark the boundaries of the regular stripe patterns, the arrow labeled $\varepsilon$ indicates the direction of detuning between the left and right QDs, and $\varepsilon_{\max}$ represents the scale of the stripe triangle. In the experiment, a fixed bias of $V_{QPC} = 0.2$ mV is applied to the QPC to enhance the QPC detection sensitivity. (b) The upper panel shows the dependence of the oscillation detuning-boundary spacing on the source–drain bias at $V_{QPC} = 0.2$ mV. The lower panel shows the dependence of the boundary spacing on $V_{QPC}$ while keeping the source–drain bias fixed at $V_{sd} =$ 2 mVand $\varepsilon_{max} = 2$ meV. (c) DQD stability diagram calculated using the theory in the main text, which describes the emission of acoustic phonons by the non-equilibrium hole fluctuations. $t_c = 8$ μeV , $T_h = 200\ mK$ and $\Gamma_{\mathrm{L/R}} \approx 90$ MHz are used for calculation. (d) Extracted oscillation as a function of detuning energy along the arrow in (a). The average spacing of 140 μeV between the adjacent tunnel peaks is extracted as the period of oscillation.

*3. Numerical Simulation of Oscillatory Current.* Using the bonding/anti-bonding state model, we describe the DQD as a system involving three states: $|\Psi_+\rangle = \mathrm{c}_1|L\rangle +$ $c_2|R\rangle, |\Psi_-\rangle = c_2|L\rangle - c_1|R\rangle$ (where $c_1 = \cos\frac{\theta}{2}, c_2 = \sin\frac{\theta}{2}, \theta = arctan\frac{2t_c}{\varepsilon}$) and $|0\rangle$.[32-33] The Hamiltonian describing the phonon effect in the DQDs is as follows[29-30]:

$$H_{\mathrm{ph}} = \sum_q \hbar\omega_q a_q^+ a_q + \hbar\lambda_{\mathrm{q}}\sigma_z(a_q^+ + a_q). \quad (1)$$

Using the Born-Markov approximation, the master equation for the DQD system can be expressed in the coordinate basis $|\Psi_+\rangle$, $|\Psi_-\rangle$ and $|0\rangle$ as:

$$\dot{\rho} = i[H_0, \rho] + U_{\mathrm{t}}\rho + U_{\mathrm{ph}}\rho$$

$$U_{\mathrm{t}} = c_1^2(\Gamma_{\mathrm{L}} D[|\Psi_+\rangle\langle 0|] + \Gamma_{\mathrm{R}} D[|0\rangle\langle\Psi_-|]) + c_2^2(\Gamma_{\mathrm{L}} D[|\Psi_-\rangle\langle 0|] + \Gamma_{\mathrm{R}} D[|0\rangle\langle\Psi_+|])$$

$$U_{\mathrm{ph}} = \Gamma_{\mathrm{e}} D[|\Psi_-\rangle\langle\Psi_+|] + \Gamma_{\mathrm{a}} D[|\Psi_-\rangle\langle\Psi_+|] \quad (2)$$

where $H_0 = \frac{\varepsilon}{2}\sigma_z + t_c\sigma_x$ is the Hamiltonian of the DQD charge states, $D[L]\rho = -\frac{1}{2}[L^+L, \rho] + L\rho L^+$ is the Lindblad superoperator, $L$ is the Lindblad operator, and $\Gamma_{\mathrm{R}}$ ($\Gamma_{\mathrm{L}}$) are the tunneling rates between the right (left) dot and the drain (source). The second term of equation (2) is the elastic tunneling term, which contributes only under charge resonance between the two dots. The third term describes the interaction between phonons and holes, leading to tunneling between $|\Psi_+\rangle$ and $|\Psi_-\rangle$. The first half of $U_{\mathrm{ph}}$ describes the phonon emission process, as shown in Fig. 3(b) while the latter half of $U_{\mathrm{ph}}$ describes the phonon absorption process. $\Gamma_{\mathrm{e}} = 2\pi(n_q + 1)\rho(\omega)$ is the emission coefficient proportional to $n_q + 1$, while $\Gamma_{\mathrm{a}}$ is the absorption coefficient proportional to the average phonon number $n_q = \left(e^{\frac{\hbar\omega_\lambda}{k_B T_h}} - 1\right)^{-1}$, which approximates to zero and can be ignored when $T_h < 1$ K.

Given the tunneling rates between the left (right) quantum dot and the source (drain) $\Gamma_{\mathrm{L/R}}$, as well as the interdot coupling $t_c$, the steady-state probabilities of the three states $p_+$, $p_-$ and $p_0$, can be obtained once the hole–phonon coupling strength $\rho(\epsilon)$ of the system is determined.

According to previous studies, the relative strengths of the piezoelectric coupling and deformation-potential coupling of phonons in the system are determined by the intrinsic material parameters[31]. In GaAs systems, piezoelectric acoustic interaction dominates the strength of hole-phonon coupling, while the deformation potential accounts for only about 0.1% of the total contribution[31-32]. Under the assumption of single-particle gaussian wave functions and taking piezoelectric coupling into account, a concrete expression for the hole–phonon coupling strength can be derived[30-31].

$$\rho(\epsilon) = bsin^2\theta\left(1 - \frac{\delta_\epsilon}{2\pi\epsilon}\sin\left(\frac{2\pi\epsilon}{\delta_\epsilon}\right)\right)e^{-\frac{2\pi\epsilon}{\delta_c}} \tag{3}$$

where $\delta_\epsilon = hc/d$ denotes the energy of an individual phonon., and $b = \frac{e^2h_{14}^2}{2\pi^2\rho_M\hbar c^3}$ is the coupling strength coefficient. $\rho_M = 5300\text{kg}m^{-3}$ is the mass density of the crystal, $eh_{14} = 1.38 \times 10^9\text{eV}m^{-1}$ is the piezoelectric constant, $c = 5000m/s$ is the average sound velocity in GaAs. $e^{-\frac{2\pi\epsilon}{\delta_c}}$ is a decay term, where $\delta_c$ denotes the cutoff detuning associated with the effective cutoff frequency of the electron–phonon interaction, originating from the finite spatial extent of the electronic wave functions in the quantum dots.

The transport current is then given by[30]:

$$I = e\Gamma_R \cdot \text{Tr}(\rho|R\rangle\langle R|) = e\Gamma_R(c_1^2p_+ + c_2^2p_-), \tag{4}$$

This current can be differentiated to obtain the charge-sensing current, which was measured in our experiment. Finally, the calculation result (red) based on the above method is shown in Fig. 2(c), which perfectly matches our experimental data. In Fig. 2(d), we extract a one-dimensional oscillatory trace as a function of $\varepsilon$. By tuning parameters such as the interdot tunnel coupling $t_c$, the tunneling rate $\Gamma_L$ *and* $\Gamma_R$, and the phonon temperature $T_h$, and repeatedly comparing the simulated oscillation peaks with the experimental data, we obtained a unique parameter set that provides the best overall agreement with the measured curve. From the extracted oscillation spacing $\delta_\varepsilon = 140\mu eV$ the interdot distance is calculated to be $d = hc/\delta_\varepsilon = 147.7$ nm, which is in very good agreement with the designed value of 150 nm.

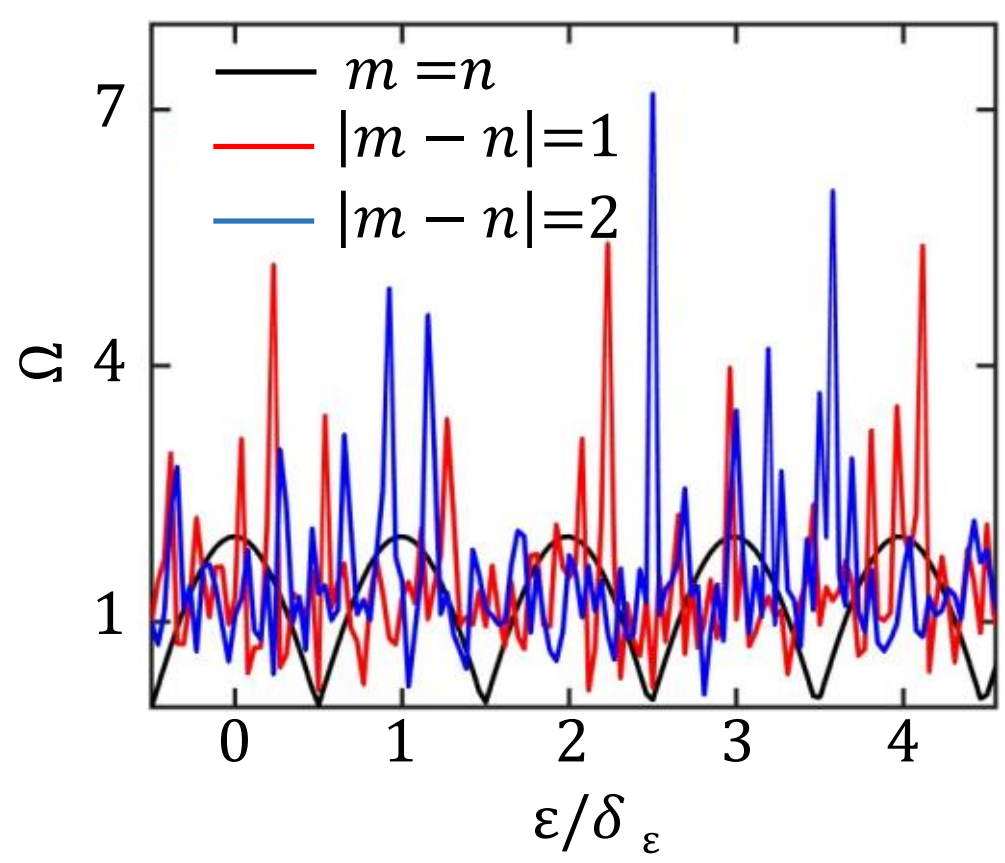

**Fig. 3.** By taking eigenfunctions of a simple harmonic-oscillator potential with different quantum numbers, we qualitatively calculate the oscillatory factor in the phonon emission rate. When $m = n$, the black curve exhibits pronounced oscillations. In contrast, for $| m - n |= 1$ and $| m - n |= 2$, the red and blue curves fluctuate around an average value close to 1, with noise-like behavior interspersed with sharp spikes.

*4. Qualitative Analysis of the Oscillatory Factors*. As shown in Fig. 1(b), measurements are performed over a larger parameter space. Within the same figure, clear oscillations are observed only in the vicinity of the $(n+1,m)$–$(n,m+1)$ tunneling region. In the neighboring $(n+1,m-1)$–$(n,m)$ and $(n+1,m+1)$–$(n,m+2)$ regions, only a few very faint lines can be discerned, while no oscillatory features are observed in more distant regions. We provide a qualitative analysis of the origin of the hole–phonon interaction being observable only in specific charge configurations.

Under the limiting conditions $\epsilon \gg \Gamma_L, \Gamma_R$ and $T_h \to 0$, the current $I$ is approximately proportional to the phonon emission rate $\Gamma_e$, indicating that the oscillatory behavior of the current originates from the oscillatory terms contained in $\Gamma_e$.

From the Fermi-golden rule[24], emission rate can be calculated as

$$\Gamma_{\mathrm{e}} = \frac{2}{(2\pi)^3\hbar}\int d^3\boldsymbol{q}\left(n_q + \frac{1}{2} \pm \frac{1}{2}\right) \times |M|^2\delta\big(\Delta - \hbar\omega_\lambda(\boldsymbol{q})\big) \tag{5}$$

where $\boldsymbol{q}$ is the phonon quasi momentum, $n_q$ the average number of phonons and $\Delta$ the energy difference between the two hole states. The relevant matrix can be written as

$$|M|^2 = \left|\langle\Psi_+|H_Q|\Psi_-\rangle\right|^2 = \left|\mathrm{c}_1 c_2\big(\langle L|e^{i\boldsymbol{qr}}|L\rangle - \langle R|e^{i\boldsymbol{qr}}|R\rangle\big)\right|^2 \tag{6}$$

including the hole-phonon interaction Hamiltonian $H_Q = \sum_q \hbar\lambda_{\mathrm{q}}\sigma_z(a_q^+ + a_q)$ and the phase term $e^{i\boldsymbol{qr}}$. As the gate architecture results in a pair of wave functions in exponential decay with $\mathbf{r}$ and well-separated by distance vector $\boldsymbol{d}$, we have $\boldsymbol{q}$ parallel with $\boldsymbol{d}$, then $c_1^2\langle 2|H_Q|1\rangle + c_2^2\langle 1|H_Q|2\rangle$ can be wiped out from the relevant matrix.

When the approximate condition is satisfied such that the wave functions in the left and right dots are similar, i.e.,

$$\Psi_{\mathrm{L}}(\boldsymbol{r}) \approx \Psi_{\mathrm{R}}(\boldsymbol{r}+\boldsymbol{d}) \tag{7}$$

In this case, one obtains $\langle R|e^{i\boldsymbol{qr}}|R\rangle = e^{i\boldsymbol{qd}}\langle L|e^{i\boldsymbol{qr}}|L\rangle$. Substituting this relation into Eq. (5) yields

$$|M|^2 = \left|\mathrm{c}_1 c_2\langle L|e^{i\boldsymbol{qr}}|L\rangle\big(1 - e^{i\boldsymbol{qd}}\big)\right|^2 \propto \big(1 - \cos(\Delta\varphi)\big) \tag{8}$$

where $\Delta\varphi = qd$. $q = \frac{2\pi\epsilon}{hc}$. In other words, the phonon emission rate exhibits oscillatory behavior as a function of $\varepsilon$, with a periodicity of $\delta_\varepsilon = hc/d$.

The validity of this model relies on the approximate condition that the tunneling hole has similar wave functions in the left and right dots, as expressed by Eq. (6). In the neighboring or more distant regions, this condition is no longer satisfied due to changes in the charge occupation of the dots. In this case, we assume that the wave functions in the left and right dots satisfy the following relation:

$$\Psi_{\mathrm{L}}(\boldsymbol{r}) = \Psi_{\mathrm{R}}(\boldsymbol{r}+\boldsymbol{d})\chi(\boldsymbol{r}+\boldsymbol{d}) \tag{9}$$

where $\chi$ denotes the transformation function of the wave functions between the left and right dots. Substituting Eq. (8) into Eq. (5) yields

$$|M|^2 = \left| c_1 c_2 \langle L|e^{i\boldsymbol{qr}}|L\rangle \left(1 - \frac{\langle L|\chi(\boldsymbol{r})e^{i\boldsymbol{qr}}|L\rangle}{\langle L|e^{i\boldsymbol{qr}}|L\rangle} e^{i\boldsymbol{qd}}\right)\right|^2 \propto$$

$$\Omega = \left|\left(1 - \frac{\langle L|\chi(\boldsymbol{r})e^{i\boldsymbol{qr}}|L\rangle}{\langle L|e^{i\boldsymbol{qr}}|L\rangle} e^{i\boldsymbol{qd}}\right)\right|^2 \tag{10}$$

We denote the oscillatory factor by $\Omega$ and perform a qualitative analysis by adopting the harmonic-oscillator potential and its wave-function solutions. The quantum numbers of the wave functions in the left and right dots are taken to be $n$ and $m$, respectively. As shown in Fig. 3, for the case $m = n$, $\Omega$ exhibits clear periodic oscillations with a period of $\delta_\varepsilon$. In contrast, for $|m-n| = 1$ and $|m-n| = 2$, $\Omega$ fluctuates around 1 with superimposed sharp spikes, indicating that the oscillatory behavior is strongly suppressed. Therefore, the observation of hole–phonon oscillatory behavior requires the similarity of the wave functions in the left and right dots.

Although oscillatory behavior may also be expected for other charge transitions such as $(m \pm 1, n \pm 1)$in principle, the actual wave-function symmetry in the device can be strongly affected by changes in the confinement potential induced by gate tuning. In experiments, tuning the charge occupation not only shifts the dot energy levels, but also modifies the shape of the confinement potential, leading to asymmetric wave functions between the two dots. As a result, the symmetry condition required for the oscillatory behavior may no longer be satisfied in the $(m \pm 1, n \pm 1)$ charge configurations, and the oscillatory stripe patterns become unobservable. Therefore, for a fixed device configuration, oscillatory stripe patterns can only be measured within a specific charge-configuration region.

*5. Summary.* We have experimentally and theoretically investigated hole–phonon interactions in a gate-defined GaAs double quantum dot integrated with a quantum point contact charge sensor. Our results show that phonon emission, with the phonon energy supplied by the applied source–drain bias, is the physical mechanism responsible for the observed inelastic interdot tunneling. Beyond identifying this phonon-assisted transport process, we find that the emergence of the oscillatory features strongly depends on the symmetry of the hole wave functions and the charge configuration of the double quantum dot. These results provide deeper insight into carrier–phonon interactions in semiconductor nanostructures and offer useful guidance for the design of quantum-dot-based devices where phonon effects play an important role[35]. In particular, tailored device structures may be engineered either to suppress phonon coupling and reduce phonon-induced decoherence in qubits[37-38], or to exploit controlled phonon coupling to realize phonon-mediated qubit interactions[39].

*Acknowledgments.* This work was supported by the National Natural Science Foundation of China (Grant Nos. 12474490 and 12574552), Quantum Science and Technology National Science and Technology Major Project (Grant No. 2021ZD0302300). This work was partially carried out at the USTC Center for Micro and Nanoscale Research and Fabrication.